\documentclass[twocolumn,eqsecnum,prb]{revtex4}
\usepackage{amsfonts}
\usepackage[T1]{fontenc}
\usepackage{amsmath,amsbsy,amssymb,graphicx}
\usepackage{times}

\def\beginABC{\begin{subequations}}
\def\endABC{\end{subequations}}

\let\mathbf=\boldsymbol

\begin{document}

\title{{\Large Quantum Hall Effects in Silicene }}
\author{Motohiko Ezawa}
\affiliation{Department of Applied Physics, University of Tokyo, Hongo 7-3-1, 113-8656,
Japan }

\begin{abstract}
We investigate quantum Hall effects in silicene by applying electric field $E_z$ parallel to magnetic field. 
Silicene is a monolayer of silicon atoms forming a two-dimensional honeycomb
lattice, and shares almost every remarkable property with graphene. A new
feature is its buckled structure, due to which the band structure can be controlled externally by changing $E_z$.
The low energy physics of silicene is described by massive Dirac fermions, 
where the mass is a function of $E_z$ and becomes zero at the critical field $E_{\text{cr}}$.
We show that there are no zero energy states due to the Dirac mass term except at the critical electric field $E_{\text{cr}}$.
Furthermore it is shown that the 4-fold degenerate zero-energy states are completely 
resolved even without considering Coulomb interactions. These features are highly contrasted with
those in graphene, demonstrating that silicene has a richer structure.
The prominent feature is that, by applying the electric field, we can control the valley degeneracy. 
As a function of $E_z$, 
Hall plateaux appear at the filling factors $\nu =0,\pm 1,\pm 2,\pm 3,\cdots $  
except for the points where level crossings occur.
\end{abstract}

\maketitle


\section{Introduction}

Electrons in graphene may be viewed as massless Dirac fermions in the
2-dimensional space\cite{GrapheneRMP}. The study of graphene started with
revealing an unusual quantum Hall (QH) effect under magnetic field\cite%
{Graphene,Graphene2,Graphene3,Graphene4,Alicea,Nomura,Goerbig,GoerbigRMP},
where Hall plateaux develop at a series of filling factors $\nu =\pm 2,\pm
4,\pm 6,\cdots $ under weak magnetic field, reflecting the existence of the
zero-energy states and the 4-fold degeneracy associated with the spin
symmetry and the valley symmetry. Though the 2-fold spin degeneracy may be
resolved in a magnetization experiment, there is no way to access to the
valley symmetry in graphene.

\begin{figure}[t]
\centerline{\includegraphics[width=0.45\textwidth]{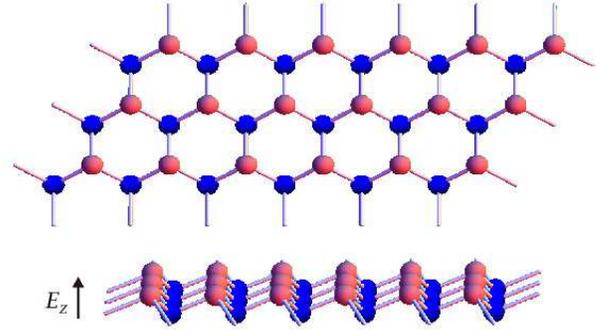}}
\caption{(Color online) Illustration of the buckled honycomb lattice of
silicene. A honeycomb lattice is distorted due to a large ionic radius of a
silicon atom and forms a buckled structure. The A and B sites form two
sublattices separated by a perpendicular distance $2\ell $. The structure
generates a staggered sublattice potential in the electric field $E_{z}$,
which leads to various intriguing pheneomena.}
\label{FigBuckl}
\end{figure}

Recently silicene, a monolayer of silicon atoms forming a two-dimensional
honeycomb lattice, has been synthesized\cite{Lalmi,Padova,Aufray} and
attracts much attention\cite{Guzman,LiuPRL,SiliceneE}. Silicene shares
almost every remarkable property with graphene. Indeed, it has Dirac cones
akin to graphene. It has additionally a salient feature, that is a buckled
structure\cite{Guzman,LiuPRL} owing to a large ionic radius of silicon (Fig.%
\ref{FigBuckl}). Consequently, silicene has a relatively large spin-orbit
(SO) gap of $1.55$meV, which provides a mass to Dirac electrons.
Furthermore, we may control experimentally the mass by applying the electric
field $E_{z}$ perpendicular to the silicene sheet. As $|E_{z}|$ increases,
the Dirac mass decreases linearly, and vanishes at the critical point $%
|E_{z}|=E_{\text{cr}}$, and then increases linearly\cite{SiliceneE}.

In this paper we analyze the QH effect in silicene, which is the first step
of exploring the intrinsic properties of silicene as in the case of
graphene. The Hamiltonian contains the effective spin-orbit coupling term ($%
\varpropto \lambda _{\text{SO}}$), the Rashba spin-orbit coupling ($%
\varpropto \lambda _{\text{R}}$) and the electric field term ($\varpropto
E_{z}$) additionally to that of graphene. The Dirac mass depends on $\lambda
_{\text{SO}}$ and $E_{z}$. There are 4-fold degenerate zero-energy states in
graphene. Let us call the counterpart the would-be zero-energy states in
silicene. Their degeneracy is completely resolved except at the critical
electric field, $E_{z}=\pm E_{\text{cr}}$, due to the mass effect. Hall
plateaux develop at $\nu =0,\pm 2$ for $E_{z}=0$, $\nu =\pm 1,\pm 2$ for $%
E_{z}=\pm E_{\text{cr}}$, and $\nu =0,\pm 1,\pm 2$ elsewhere. Hence the
critical electric field $E_{\text{cr}}$ can be experimentally measurable.
Such a measurement will be important since it has been argued\cite{SiliceneE}
that a transition occurs between topological and band insulators at this
critical field without magnetic field. Our finding will be the first case in
which we can control the Landau level by applying external electric field
parallel to the magnetic field.

This paper is composed as follows. In Section \ref{SecDirac} we calculate
the energy levels in the silicene system under magnetic field as a function
of $E_{z}$. This allows us to determine the series of filling factors at
each $E_{z}$. In Section \ref{SecSpectrum} we elucidate fully the physical
meaning of the energy spectrum. When we switch off the Rashba coupling ($%
\lambda _{\text{R}}=0$), the Hamiltonian is diagonalized analytically and
the physical picture becomes manifest. Two successive Landau levels are
mixed by the intrinsic Zeeman effect acting between the A and B sublattaices
of a honeycomb lattice just as in the case of graphene\cite{EzawaJPSJ}. The
underlying symmetry is the supersymmetry\cite{EzawaPLA} even in the presence
of the Dirac mass term. Each energy level is 2-fold degenerate in general.
On the other hand, as we have mentioned, the four states in the would-be
zero energy states are not degenerate: Namely, each of these states contains
electrons with a definite spin polarization either from the K or K' point.
The effect of the Rashba coupling ($\lambda _{\text{R}}\neq 0$) is to remove
the 2-fold valley degeneracy by modifying the energy level in the order of $%
\lambda _{\text{R}}$. Furthermore, it mixes an up-spin state and a down-spin
state at a certain crossing point and to turn it into an anticrossing point.
In conclusion, all the degeneracy is removed except for the points at which
two levels cross.

\section{Low-Energy Dirac Theory}

\label{SecDirac}

Silicene consists of a honeycomb lattice of silicon atoms with two
sublattices made of A sites and B sites. The states near the Fermi energy
are $\pi $ orbitals residing near the K and K' points at opposite corners of
the hexagonal Brillouin zone. (We also call them the K$_{\eta }$ points with 
$\eta =\pm $.) We take a silicene sheet on the $xy$-plane, and apply the
electric field $E_{z}$ perpendicular to the plane. Due to the buckled
structure the two sublattice planes are separated by a distance, which we
denote by $2\ell $ with $\ell =0.23$\AA\ , as illustrated in Fig.\ref%
{FigBuckl}. It generates a staggered sublattice potential $\varpropto 2\ell
E_{z}$ between silicon atoms at A sites and B sites.

\begin{figure}[t]
\centerline{\includegraphics[width=0.45\textwidth]{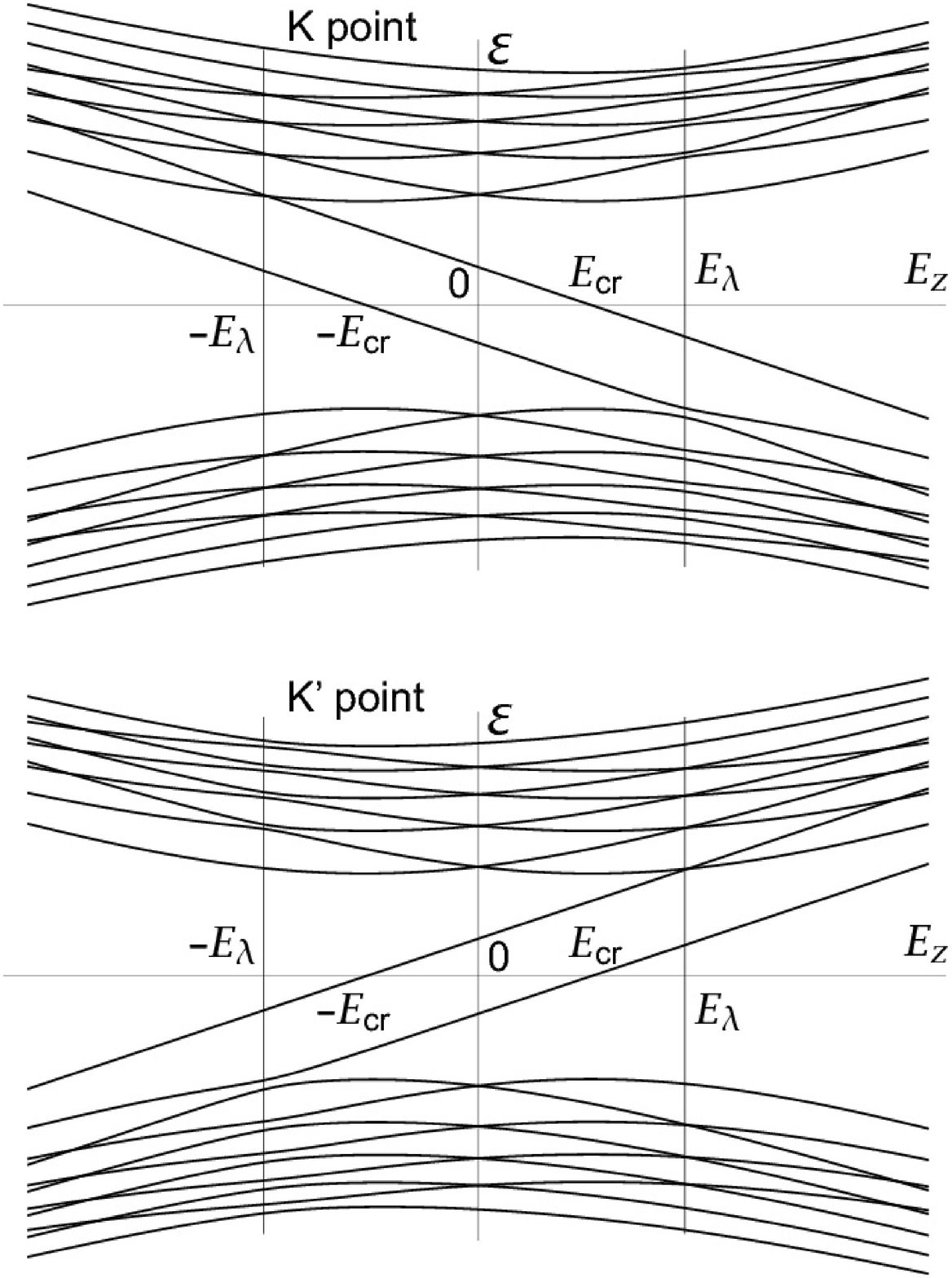}}
\caption{Energy levels as a function of $E_{z}$ for $\protect%
\lambda _{R}\neq 0$. There exist two levels which yield the zero energy at $%
E_{z}=\pm E_{\text{cr}}$ in the K valley, and also in the K' valley. (We
call them the would-be zero-energy states.) Level crossing occurs at $%
E_{z}=nE_{\protect\lambda }$, $n=0,\pm 1,\pm 2,\cdots $, except at $E_{z}=%
\protect\eta E_{\protect\lambda }$ in the K$_{\protect\eta }$ valley where
anticrossing takes over. See (\protect\ref{CrossE}) with respect to $E_{%
\protect\lambda }$.}
\label{FigLLr}
\end{figure}

We analyze the physics of electrons near the Fermi energy more. We employ
the low-energy Dirac theory, which has been proved to be useful in the study
of graphene\cite{Semenoff1,Semenoff2,Ajiki,Ando}. The low-energy effective
Hamiltonian around the K$_{\eta }$ point reads as\cite{LiuPRL}%
\begin{equation}
H_{\eta }=\hbar v_{\text{F}}\left( k_{x}\tau _{x}-\eta k_{y}\tau _{y}\right)
+\eta \tau _{z}h_{11}+\ell E_{z}\tau _{z},  \label{DiracHamil}
\end{equation}%
with%
\begin{equation}
h_{11}=-\lambda _{\text{SO}}\sigma _{z}-a\lambda _{\text{R}}\left(
k_{y}\sigma _{x}-k_{x}\sigma _{y}\right) ,
\end{equation}%
where $\tau _{a}$ is the Pauli matrix of the sublattice, $v_{\text{F}}=\frac{%
\sqrt{3}}{2}at=5.5\times 10^{5}$m/s is the Fermi velocity, $a=3.86$\AA\ is
the lattice constant, $\lambda _{\text{SO}}=3.9$meV is the effective
spin-orbit coupling, and $\lambda _{\text{R}}=0.7$meV is the intrinsic
Rashba spin-orbit coupling. The two Hamiltonians $H_{+}$ and $H_{-}$ are
related through the time-reversal operation. See the Appendix with respect
to the tight-binding model and the low-energy Dirac theory in silicene.

\begin{figure}[t]
\centerline{\includegraphics[width=0.45\textwidth]{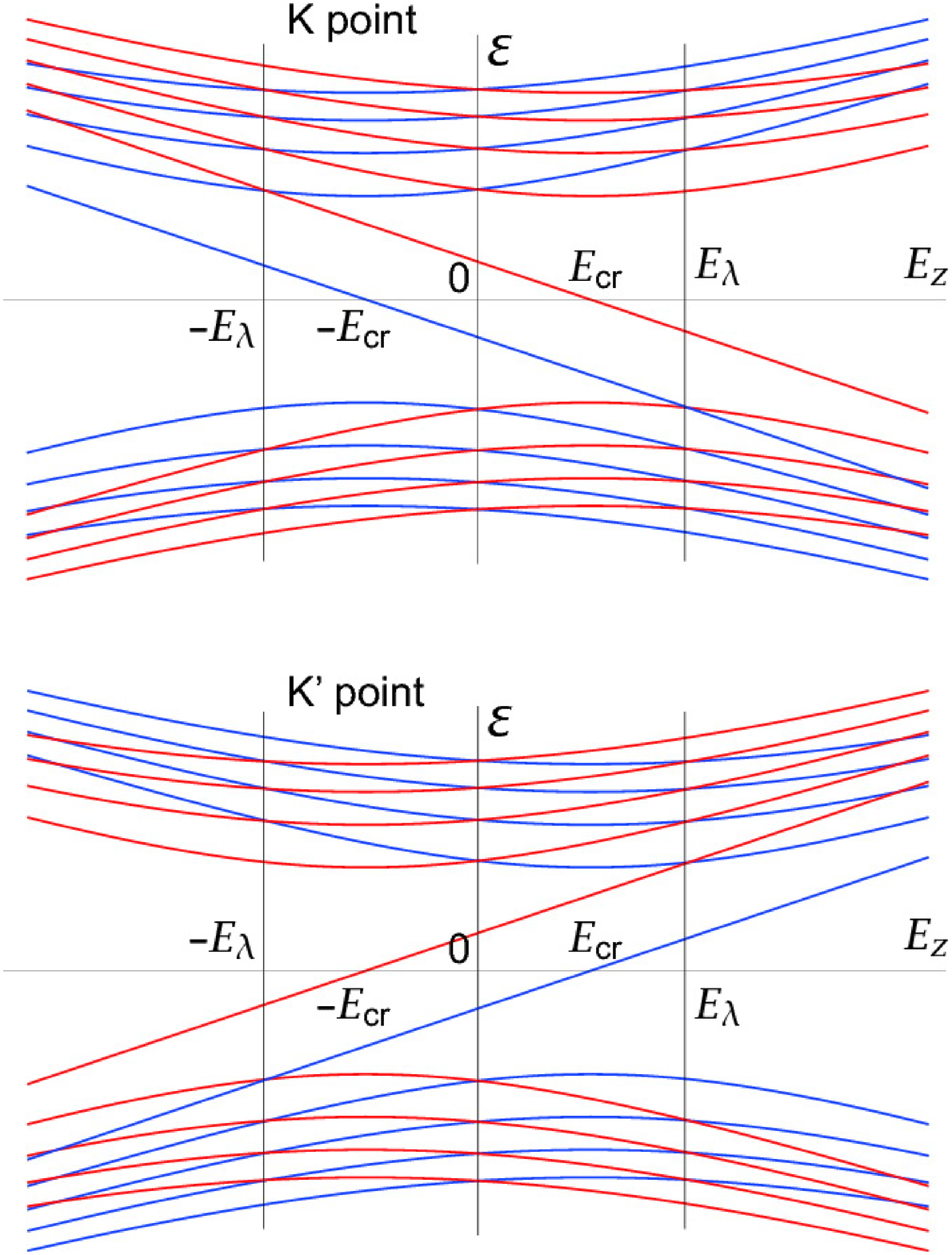}}
\caption{(Color online) Energy levels as a function of $E_{z}$ for $\protect%
\lambda _{R}=0$. All electrons are spin-up (red curves) or spin-down (blue
curves) polarized. There exist two levels which yield the zero energy at $%
E_{z}=\pm E_{\text{cr}}$ in the K valley, and also in the K' valley. Level
crossing occurs at $E_{z}=nE_{\protect\lambda }$, $n=0,\pm 1,\pm 2,\cdots $.
There exists no anticrossing. Compare the spectrum with the one in Fig.%
\protect\ref{FigLLr}.}
\label{FigLL}
\end{figure}

The energy spectrum is readily derived from (\ref{DiracHamil}) as%
\begin{equation}
\mathcal{E}_{\eta s_{z}}=\pm \sqrt{\hbar ^{2}v_{\text{F}}^{2}k^{2}+\left(
\ell E_{z}-\eta s_{z}\sqrt{\lambda _{\text{SO}}^{2}+a^{2}\lambda _{\text{R}%
}^{2}k^{2}}\right) ^{2}}.  \label{gapDirac}
\end{equation}%
The gap is given by $2|\Delta _{\eta s_{z}}\left( E_{z}\right) |$ with%
\begin{equation}
\Delta _{\eta s_{z}}\left( E_{z}\right) =-\eta s_{z}\lambda _{\text{SO}%
}+\ell E_{z}.  \label{gapDiracX}
\end{equation}%
As $|E_{z}|$ increases, the gap decreases linearly, and vanishes at the
critical point $|E_{z}|=E_{\text{cr}}$ with%
\begin{equation}
E_{\text{cr}}=\eta s_{z}\lambda _{\text{SO}}/\ell =17\text{meV/\AA },
\end{equation}%
and then increases linearly. Note that the spectrum depends on the spin
index $s_{z}$ and the valley index $\eta $ only in the combination $\eta
s_{z}$.

We apply a homogeneous magnetic field $\mathbf{B}=\mathbf{\nabla }\times 
\mathbf{A}=\left( 0,0,-B\right) $ with $B>0$ along the $z$ axis to silicene.
By making the minimal substitution, the Hamiltonian is given by%
\begin{equation}
H_{\eta }=\hbar v_{\text{F}}\left( P_{x}\tau _{x}-\eta P_{y}\tau _{y}\right)
+\eta \tau _{z}h_{11}+\ell E_{z}\tau _{z}
\end{equation}%
with the covariant momentum $P_{i}\equiv k_{i}+eA_{i}$. We introduce a pair
of Landau-level ladder operators, 
\begin{equation}
\hat{a}=\frac{\ell _{B}(P_{x}+iP_{y})}{\sqrt{2}\hbar },\quad \hat{a}%
^{\dagger }=\frac{\ell _{B}(P_{x}-iP_{y})}{\sqrt{2}\hbar },  \label{G-OperaA}
\end{equation}%
satisfying $[\hat{a},\hat{a}^{\dag }]=1$, where $\ell _{B}=\sqrt{\hbar /eB}$
is the magnetic length. In the basis $\left\{ \psi _{A\uparrow },\psi
_{B\uparrow },\psi _{A\downarrow },\psi _{B\downarrow }\right\} ^{t}$, the
Hamiltonian $H_{\eta }$ reads%
\begin{equation}
\left( 
\begin{array}{cccc}
\Delta _{+}\left( E_{z}\right) & \hbar \omega _{\text{c}}\hat{a} & i\frac{%
\sqrt{2}\hbar a\lambda _{\text{R}}}{\ell _{B}}\hat{a}^{\dagger } & 0 \\ 
\hbar \omega _{\text{c}}\hat{a}^{\dagger } & -\Delta _{+}\left( E_{z}\right)
& 0 & -i\frac{\sqrt{2}\hbar a\lambda _{\text{R}}}{\ell _{B}}\hat{a}^{\dagger
} \\ 
-i\frac{\sqrt{2}\hbar a\lambda _{\text{R}}}{\ell _{B}}\hat{a} & 0 & \Delta
_{-}\left( E_{z}\right) & \hbar \omega _{\text{c}}\hat{a} \\ 
0 & i\frac{\sqrt{2}\hbar a\lambda _{\text{R}}}{\ell _{B}}\hat{a} & \hbar
\omega _{\text{c}}\hat{a}^{\dagger } & -\Delta _{-}\left( E_{z}\right)%
\end{array}%
\right) ,  \label{HamilBK}
\end{equation}%
at the K point, with $\omega _{\text{c}}=\sqrt{2}\hbar v_{\text{F}}/\ell
_{B} $, and%
\begin{equation}
\left( 
\begin{array}{cccc}
\Delta _{-}\left( E_{z}\right) & \hbar \omega _{\text{c}}\hat{a}^{\dagger }
& -i\frac{\sqrt{2}\hbar a\lambda _{\text{R}}}{\ell _{B}}\hat{a}^{\dagger } & 
0 \\ 
\hbar \omega _{\text{c}}\hat{a} & -\Delta _{-}\left( E_{z}\right) & 0 & i%
\frac{\sqrt{2}\hbar a\lambda _{\text{R}}}{\ell _{B}}\hat{a}^{\dagger } \\ 
i\frac{\sqrt{2}\hbar a\lambda _{\text{R}}}{\ell _{B}}\hat{a} & 0 & \Delta
_{+}\left( E_{z}\right) & \hbar \omega _{\text{c}}\hat{a}^{\dagger } \\ 
0 & -i\frac{\sqrt{2}\hbar a\lambda _{\text{R}}}{\ell _{B}}\hat{a} & \hbar
\omega _{\text{c}}\hat{a} & -\Delta _{+}\left( E_{z}\right)%
\end{array}%
\right) ,  \label{HamilBKp}
\end{equation}%
at the K' point. Here the index $+$ of $\Delta _{+}\left( E_{z}\right) $
implies $\eta s_{z}=+$, that is, either $(s_{z},\eta )=(+,+)$ or $%
(s_{z},\eta )=(-,-)$.

Inspecting the Hamiltonian $H_{\eta }$ we see that the eigenstate is of the
form%
\begin{equation}
\Psi _{+}^{N}=\left( u_{A\uparrow }^{N}\left\vert N\right\rangle
,u_{B\uparrow }^{N+1}\left\vert N+1\right\rangle ,u_{A\downarrow
}^{N-1}\left\vert N-1\right\rangle ,u_{B\downarrow }^{N}\left\vert
N\right\rangle \right) ^{t},  \label{EigenStateS}
\end{equation}%
and%
\begin{equation}
\Psi _{-}^{N}=\left( v_{A\uparrow }^{N+1}\left\vert N+1\right\rangle
,v_{B\uparrow }^{N}\left\vert N\right\rangle ,v_{A\downarrow }^{N}\left\vert
N\right\rangle ,v_{B\downarrow }^{N-1}\left\vert N-1\right\rangle \right)
^{t},
\end{equation}%
with%
\begin{equation}
\left\vert N\right\rangle =\frac{1}{\sqrt{N!}}a^{\dagger N}\left\vert
0\right\rangle ,
\end{equation}%
which represents the state in the $N$th Landau level. It is notable that an
energy eigenstate is a mixture of four states coming from three different
Landau levels in general. The state (\ref{EigenStateS}) is defined for $%
N\geq 1$. There exists two more states corresponding to $N=-1$ and $N=0$ in (%
\ref{EigenStateS}),%
\begin{align}
\Psi _{+}^{0\uparrow }& =\left( 0,\left\vert 0\right\rangle ,0,0\right) ^{t},
\label{EigenStateS0u} \\
\Psi _{+}^{0\downarrow }& =\left( u_{A\uparrow }^{0}\left\vert
0\right\rangle ,u_{B\uparrow }^{1}\left\vert 1\right\rangle
,0,u_{B\downarrow }^{0}\left\vert 0\right\rangle \right) ^{t},
\label{EigenStateS0d}
\end{align}%
and%
\begin{align}
\Psi _{-}^{0\uparrow }& =\left( \left\vert 0\right\rangle ,0,0,0\right) ^{t},
\\
\Psi _{-}^{0\downarrow }& =\left( v_{A\uparrow }^{1}\left\vert
1\right\rangle ,v_{B\uparrow }^{0}\left\vert 0\right\rangle ,v_{A\downarrow
}^{0}\left\vert 0\right\rangle ,0\right) ^{t}.
\end{align}%
They exhaust all the eigenstates of the Hamiltonian $H_{\eta }$. Applying
the Hamiltonian $H_{\eta }$ to these states, we can determine numerically
the energy spectrum as a function of $E_{z}$, which we display in Fig.\ref%
{FigLLr}. All energy levels are nondegenerate except for certain isolated
values of $E_{z}$ including $E_{z}=0$. In conclusion, the 4-fold degeneracy
in graphene is completely resolved except for these isolated points, which
yields the QH plateaux at $\nu =0,\pm 1,\pm 2,\pm 3,\cdots $. It is
remarkable that the degeneracy associated with the K-K' symmetry is resolved
in silicene.

It is instructive to set $\lambda _{\text{R}}=0$ in the Hamiltonians (\ref%
{HamilBK}) and (\ref{HamilBKp}), since the Rashba coupling $\lambda _{\text{R%
}}$ is a small parameter compared with the others. The Hamiltonian becomes
block diagonal, where the spin $s_{z}$ is a good quantum number. Namely,
each energy level has a definite spin polarization at the K and K' point
separately. The resultant Hamiltonians are exactly solvable. We give the
energy spectrum as a function of $E_{z}$ in Fig.\ref{FigLL}. By comparing
the spectrum for $\lambda _{\text{R}}\neq 0$ and $\lambda _{\text{R}}=0$, it
is found that the spectra look almost the same between them except for a
certain value of $E_{z}$, that is, $E_{z}=\eta E_{\lambda }$ in the K$_{\eta
}$ valley. At this value a simple level crossing occurs for $\lambda _{\text{%
R}}=0$, but an anticrossing takes place for $\lambda _{\text{R}}\neq 0$.
Though the spectra look quite the same, there exists actually a correction
of the energy of the order $\lambda _{\text{R}}$ in the system with $\lambda
_{\text{R}}\neq 0$, which removes the degeneracy from all levels except for
level crossing points.

\section{Energy Spectrum}

\label{SecSpectrum}

We explore the energy spectrum in detail. The silicene Hamiltonian (\ref%
{BasicHamil}) is reduced to the graphene Hamiltonian if we set $\lambda _{%
\text{SO}}=\lambda _{\text{R}}=0$ and $\ell =0$, where there exists the
4-fold degenerate zero-energy states. Let us call them the would-be zero
modes in silicene. We wish to study how the degeneracy is removed. We
discuss the states (\ref{EigenStateS0u}), (\ref{EigenStateS0d}), and (\ref%
{EigenStateS}) in this order.

First of all, we note that the leading order of the Hamiltonian $H_{\eta }$
takes a simple form in the limit $E_{z}\rightarrow \pm \infty $,%
\begin{equation}
H_{\eta }=\ell E_{z}\,[\text{diag.}(1,-1,1,-1)+O(1/E_{z})].
\end{equation}%
Consequently, there are four eigenstates in this limit such that\beginABC%
\begin{align}
\text{(i)}& \text{\quad }u_{A\uparrow }^{N}=1,\quad u_{B\uparrow
}^{N+1}=u_{A\downarrow }^{N-1}=u_{B\downarrow }^{N}=0, \\
\text{(ii)}& \text{\quad }u_{B\uparrow }^{N+1}=1,\quad u_{A\uparrow
}^{N}=u_{A\downarrow }^{N-1}=u_{B\downarrow }^{N}=0, \\
\text{(iii)}& \text{\quad }u_{A\downarrow }^{N-1}=1,\quad u_{A\uparrow
}^{N}=u_{B\uparrow }^{N+1}=u_{B\downarrow }^{N}=0, \\
\text{(iv)}& \text{\quad }u_{B\downarrow }^{N}=1,\quad u_{A\uparrow
}^{N}=u_{B\uparrow }^{N+1}=u_{A\downarrow }^{N-1}=0.
\end{align}%
\endABC It turns out that we can label each energy level by the two
asymptotic states in the limit $E_{z}\rightarrow \pm \infty $. For instance,
the label [$N$A$\uparrow ,N$B$\downarrow $] implies that we have $%
u_{A\uparrow }^{N}=1$ as $E_{z}\rightarrow -\infty $ and $u_{B\downarrow
}^{N}=1$ as $E_{z}\rightarrow +\infty $.

\subsection{Would-be Zero-Energy States}

\begin{figure}[t]
\centerline{\includegraphics[width=0.45\textwidth]{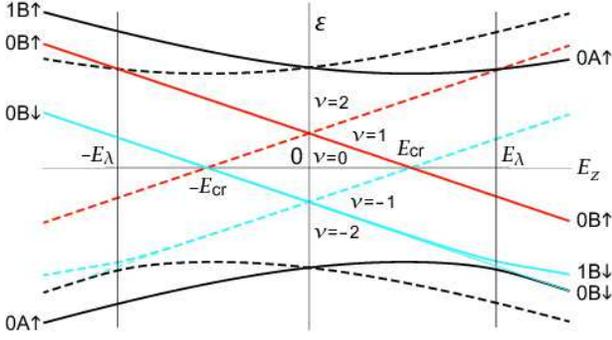}}
\caption{(Color online) Energy levels as a function of $E_{z}$ for the state 
$\Psi _{\protect\eta }^{0\uparrow }$ and $\Psi _{\protect\eta }^{0\downarrow
}$. There are four levels (solid) from the K point and four levels (dotted)
for the K' point. The red lines are the would-be zero mode made entirely of
up spins. On the other hand, the blue thin lines are the would-be zero mode
made of entirely of down spin if $\protect\lambda _{R}=0$. They anticross
with other levels with up spin at $E_{z}=\pm E_{\protect\lambda}$, and
become mixed states when $\protect\lambda _{R}\neq 0$. It is remarkable that
there exists no degeneracy for $0<|E_{z}|<E_{\text{cr}}$. Namely, both the
spin and valley symmeties are explicitly broken, where the QH plateaux
appear at $\protect\nu =0,\pm 1,\pm 2$. The symbol such as $N$A$\uparrow $
in the left (right) column indicates that the state is dominated by up-spin
electrons at the A site coming from the $N$th Landau level as $%
E_{z}\rightarrow -\infty $ ($E_{z}\rightarrow +\infty $). }
\label{FigZeroLL}
\end{figure}

For the state (\ref{EigenStateS0u}) it is trivial to solve the eigenvalue
problem,%
\begin{equation}
H_{\eta }\Psi _{\eta }^{0\uparrow }=\varepsilon _{\eta }^{\uparrow }\Psi
_{\eta }^{0\uparrow },
\end{equation}%
where the eigenvalue is exactly given by%
\begin{equation}
\varepsilon _{\eta }^{\uparrow }=-\eta \Delta _{\eta }\left( E_{z}\right) ,
\end{equation}%
with (\ref{gapDiracX}). This is a reminiscence of the gap energy in the
system without magnetic field. In particular it implies the emergence of the
zero-energy state ($\varepsilon _{\eta }^{\uparrow }=0$) at the critical
point $E_{z}=\eta E_{\text{cr}}$. We illustrate the energy levels by red
lines (solid line from the K point and dotted line from the K' point) in Fig.%
\ref{FigZeroLL}. The label of this state is [0B$\uparrow $,0B$\uparrow $] at
the K point and [0A$\uparrow $,0A$\uparrow $] at the K' point according to
our convention.

There are three states of the type (\ref{EigenStateS0d}). It is not easy to
solve them analytically. However, when we set $\lambda _{\text{R}}=0$, the
zero-th order Hamiltonian $H_{\eta }^{(0)}$ becomes block diagonal,%
\begin{equation}
H_{\eta }^{(0)}=\left( 
\begin{array}{cc}
H_{\eta }^{\uparrow } & 0 \\ 
0 & H_{\eta }^{\downarrow }%
\end{array}%
\right) ,  \label{HamilBrockA}
\end{equation}%
with\beginABC\label{BlockEleme}%
\begin{align}
H_{+}^{s_{z}}& =\left( 
\begin{array}{cc}
\Delta _{\eta s_{z}}\left( E_{z}\right) & \hbar \omega _{\text{c}}\hat{a} \\ 
\hbar \omega _{\text{c}}\hat{a}^{\dagger } & -\Delta _{\eta s_{z}}\left(
E_{z}\right)%
\end{array}%
\right) , \\
H_{-}^{s_{z}}& =\left( 
\begin{array}{cc}
\Delta _{\eta s_{z}}\left( E_{z}\right) & \hbar \omega _{\text{c}}\hat{a}%
^{\dagger } \\ 
\hbar \omega _{\text{c}}\hat{a} & -\Delta _{\eta s_{z}}\left( E_{z}\right)%
\end{array}%
\right) .
\end{align}%
\endABC It is easy to diagonalize $H_{\eta }^{s_{z}}$ analytically, where
the spin $s_{z}$ is a good quantum number.

Now it is trivial to obtain one of the eigenstates (\ref{EigenStateS0d}) at
the K point such that $u_{A\uparrow }^{0}=u_{B\uparrow }^{1}=u_{A\downarrow
}^{0}=0$ and $u_{B\downarrow }^{0}=1$. The state is labelled by [0B$%
\downarrow $,0B$\downarrow $], as indicated by a blue thin line in Fig.\ref%
{FigZeroLL}. It leads to the emergence of the zero-energy state at the
critical point $E_{z}=-E_{\text{cr}}$. When we include the effect of $%
\lambda _{\text{R}}\neq 0$, however, an anticrossing takes place between the
level and another level at $E_{z}=E_{\lambda }$ as indicated by a blue line,
and the resulting state becomes [0B$\downarrow $,1B$\downarrow $]. We
discuss the anticrossing later.

Similarly we can treat $H_{-}^{(0)}$. The level is indicated by a blue
dotted thin line in Fig.\ref{FigZeroLL}. It leads to the emergence of the
zero-energy state at the critical point $E_{z}=E_{\text{cr}}$. As an effect
of $\lambda _{\text{R}}\neq 0$, an anticrossing takes place with another
level at $E_{z}=-E_{\lambda }$.

We have discussed the four states coming from the K and K' points. They are
the would-be zero-energy modes in silicene, which become degenerate
zero-energy states in the limit $\lambda _{\text{SO}}=\lambda _{\text{R}}=0$%
. The degeneracy is completely resolved except at the critical electric
field, $E_{z}=\pm E_{\text{cr}}$, due to the mass effect. Hall plateaux
develop at $\nu =0,\pm 2$ for $E_{z}=0$, $\nu =\pm 1,\pm 2$ for $E_{z}=\pm
E_{\text{cr}}$, and $\nu =0,\pm 1,\pm 2$ elsewhere. Hence the critical
electric field $E_{\text{cr}}$ can be determined experimentally.

There exists two more states of the type (\ref{EigenStateS0d}), which are
indicated by black curves with the label [$1$B$\uparrow $,$0$A$\uparrow $]
and [$0$A$\uparrow $,0B$\downarrow $] in Fig.\ref{FigZeroLL}. We discuss
them in the succeeding section.

\subsection{Nonzero Modes}

\begin{figure}[t]
\centerline{\includegraphics[width=0.45\textwidth]{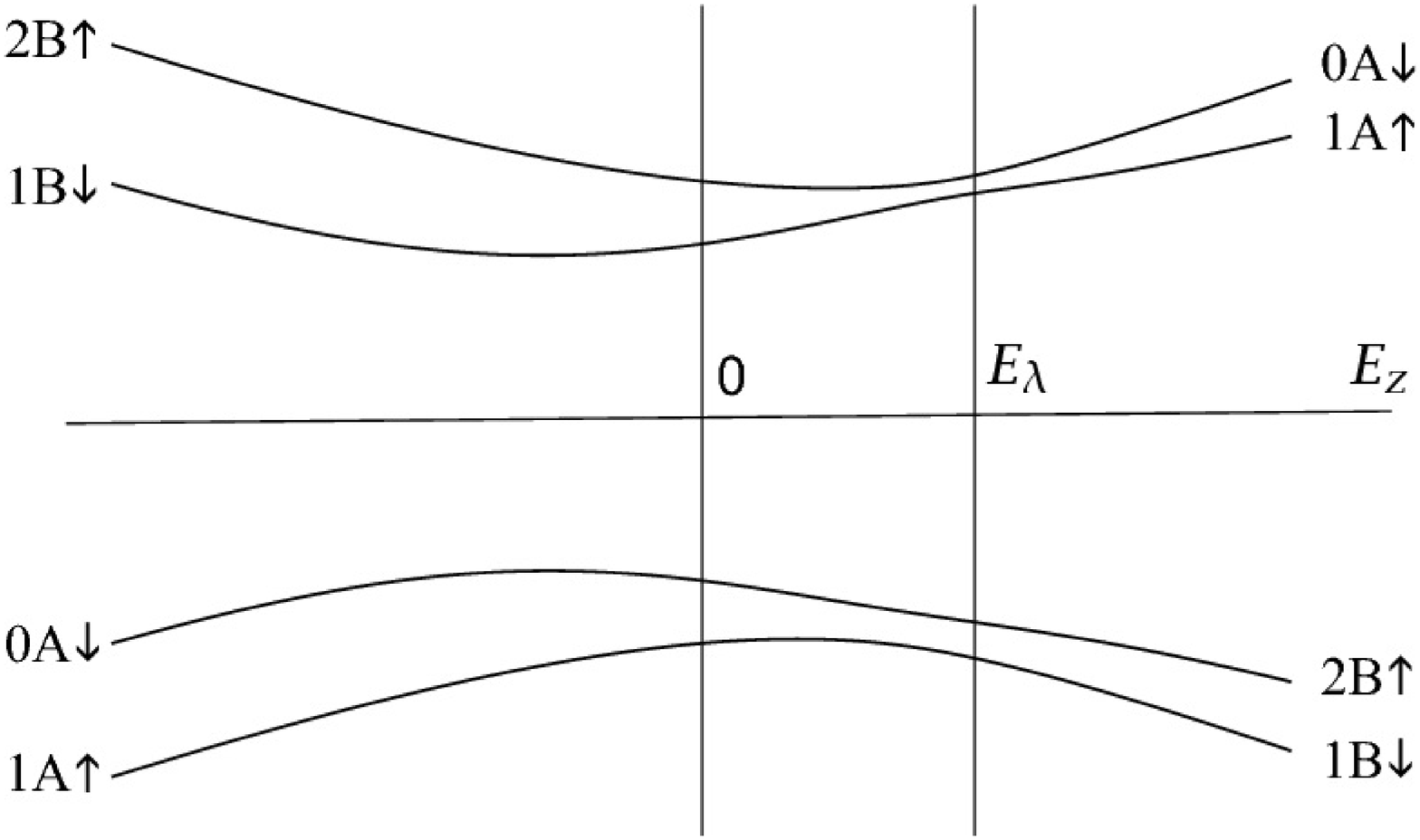}}
\caption{Energy levels as a function of $E_{z}$ for $\protect%
\lambda _{R}\neq 0$ comming from the state $\Psi _{+}^{1}$. Each\ level is
dominated by up or down-spin electrons at A or B sites in the limit $%
E_{z}\rightarrow \pm \infty $, as indicated, but they mix near the
anticrossing point ($Ez=E_{\protect\lambda }$). }
\label{FigCross}
\end{figure}

We go on to analyze the eigenstate (\ref{EigenStateS}), i.e.,%
\begin{equation}
\Psi _{+}^{N}=\left( u_{A\uparrow }^{N}\left\vert N\right\rangle
,u_{B\uparrow }^{N+1}\left\vert N+1\right\rangle ,u_{A\downarrow
}^{N-1}\left\vert N-1\right\rangle ,u_{B\downarrow }^{N}\left\vert
N\right\rangle \right) ^{t}.
\end{equation}%
There are four eigenstates for each $N$. We set $\lambda _{\text{R}}=0$,
where the Hamiltonian becomes block diagonal as in (\ref{HamilBrockA}). The
Hamiltonian (\ref{BlockEleme}) is a Dirac Hamiltonian with spin dependent
mass. The four eigenstates read as follows.

With respect to the up-spin sector, the eigenvalue problems are solved as%
\begin{equation}
H_{+}^{\uparrow }\left( 
\begin{array}{c}
u_{A\uparrow }^{N}\left\vert N\right\rangle \\ 
u_{B\uparrow }^{N+1}\left\vert N+1\right\rangle%
\end{array}%
\right) =\pm \varepsilon _{N}^{\uparrow }\left( 
\begin{array}{c}
u_{A\uparrow }^{N}\left\vert N\right\rangle \\ 
u_{B\uparrow }^{N+1}\left\vert N+1\right\rangle%
\end{array}%
\right) ,
\end{equation}%
where\beginABC\label{TypeUpSpin}%
\begin{align}
u_{A\uparrow }^{N}=& [\Delta _{+}\left( E_{z}\right) \pm \varepsilon
_{N}^{\uparrow }]/C, \\
u_{B\uparrow }^{N+1}=& \hbar \omega _{c}\sqrt{N+1}/C, \\
u_{A\downarrow }^{N-1}=& u_{B\downarrow }^{N}=0,
\end{align}%
\endABC with $C$ the normalization constant. The eigenvalue is%
\begin{equation}
\varepsilon _{N}^{\uparrow }=\varepsilon _{N+1},
\end{equation}%
with%
\begin{equation}
\varepsilon _{N}=\sqrt{\left( \hbar \omega _{c}\right) ^{2}N+\Delta _{\eta
s_{z}}^{2}\left( E_{z}\right) }.  \label{EigenZeroN}
\end{equation}%
They are labelled by [$N$A$\uparrow $,($N+1$)B$\uparrow $] and [($N+1$)B$%
\uparrow $,$N$A$\uparrow $]. See Fig.\ref{FigCross} for the case of $N=1$.

With respect to the down-spin sector, the eigenvalue problems are solved as%
\begin{equation}
H_{+}^{\downarrow }\left( 
\begin{array}{c}
u_{A\downarrow }^{N-1}\left\vert N-1\right\rangle \\ 
u_{B\downarrow }^{N}\left\vert N\right\rangle%
\end{array}%
\right) =\pm \varepsilon _{N}^{\downarrow }\left( 
\begin{array}{c}
u_{A\downarrow }^{N-1}\left\vert N-1\right\rangle \\ 
u_{B\downarrow }^{N}\left\vert N\right\rangle%
\end{array}%
\right) ,
\end{equation}%
where\beginABC\label{TypeDownSpin}%
\begin{align}
u_{A\uparrow }^{N}=& _{B\uparrow }^{N+1}=0, \\
u_{A\downarrow }^{N-1}=& [-\Delta _{-}\left( E_{z}\right) \pm \varepsilon
_{N}^{\downarrow }]/C, \\
u_{B\downarrow }^{N}=& \hbar \omega _{c}\sqrt{N}/C,
\end{align}%
\endABC with $C$ the normalization constant. The eigenvalue is%
\begin{equation}
\varepsilon _{N}^{\downarrow }=\varepsilon _{N},
\end{equation}%
with (\ref{EigenZeroN}). They are labelled by [($N-1$)A$\downarrow $,$N$B$%
\downarrow $] and [$N$B$\downarrow $,($N-1$)A$\downarrow $]. See Fig.\ref%
{FigCross} for the case of $N=1$. We can similarly discuss the K' point.

In the previous subsection we have studied $\Psi _{+}^{0}$, where we have
pointed out that there are two levels labelled by [$1$B$\uparrow $,$0$A$%
\uparrow $] and [$0$A$\uparrow $,0B$\downarrow $], which are nonzero modes
(see Fig.\ref{FigZeroLL}). The are actually the states just studied in (\ref%
{TypeUpSpin}) by choosing $N=0$.

We illustrate the energy levels in Fig.\ref{FigLL}. One level crosses with
other levels. The crossing point is given by $\varepsilon _{N}=\varepsilon
_{N^{\prime }}$ in general, which is solved as%
\begin{equation}
E_{z}=(N-N^{\prime })E_{\lambda },
\end{equation}%
with%
\begin{equation}
E_{\lambda }=\left( \hbar \omega _{c}\right) ^{2}/2\lambda _{\text{SO}}\ell .
\label{CrossE}
\end{equation}%
The spectrum with $\lambda _{\text{R}}=0$ is found to be almost identical to
that of the full theory ($\lambda _{\text{R}}\neq 0$) in Fig.\ref{FigLLr}
except for the vicinity of $E_{z}=\eta E_{\lambda }$ in the K$_{\eta }$
valley, where an anticrossing takes place.

\subsection{Landau-Level Mixing and Anticrossing}

We make a further study of the physical meaning of the energy levels. It is
a prominent feature that each energy eigenstate is a coherent superposition
of states belonging to different Landau levels. We explain the mechanism how
such a Landau-level mixing occurs by setting $\lambda _{\text{R}}=0$ and
then discuss the anticrossing by taking the effect of $\lambda _{\text{R}%
}\neq 0$.

When $\lambda _{\text{R}}=0$, the Hamiltonian is block diagonal as in (\ref%
{HamilBrockA}), where the block elements are%
\begin{equation}
H_{\eta }^{s_{z}}=\hbar \omega _{\text{c}}Q_{\eta s_{z}}+\tau _{3}\Delta
_{\eta s_{z}},
\end{equation}%
with%
\begin{equation}
Q_{+}=\left( 
\begin{array}{cc}
0 & \hat{a} \\ 
\hat{a}^{\dagger } & 0%
\end{array}%
\right) ,\quad Q_{-}=\left( 
\begin{array}{cc}
0 & \hat{a}^{\dagger } \\ 
\hat{a} & 0%
\end{array}%
\right) .  \label{SUSYQ}
\end{equation}%
To reveal the intrinsic structure of the energy spectrum, we note that%
\begin{equation}
(H_{\eta }^{s_{z}})^{2}=(\hbar \omega _{\text{c}})^{2}Q_{\eta s_{z}}Q_{\eta
s_{z}}+\Delta _{\eta s_{z}}^{2}\left( E_{z}\right) ,  \label{PauliHamilX}
\end{equation}%
and explore the Hamiltonian $H_{\text{P}}^{\pm }=Q_{\pm }Q_{\pm }$. We
explicitly discuss%
\begin{equation}
H_{\text{P}}^{+}=\left( 
\begin{array}{cc}
\hat{a}\hat{a}^{\dagger } & 0 \\ 
0 & \hat{a}^{\dagger }\hat{a}%
\end{array}%
\right) ,  \label{SUSY-Hamil}
\end{equation}%
but a similar result follows also for $H_{\text{P}}^{-}$. We set%
\begin{equation}
Q=\left( 
\begin{array}{cc}
0 & 0 \\ 
\hat{a}^{\dagger } & 0%
\end{array}%
\right) ,\qquad Q^{\dagger }=\left( 
\begin{array}{cc}
0 & \hat{a} \\ 
0 & 0%
\end{array}%
\right) .
\end{equation}%
It is easy to see%
\begin{equation}
H_{\text{P}}^{+}=\{Q,Q^{\dagger }\},\qquad \left[ H_{\eta }^{+},Q\right] =0.
\end{equation}%
It defines a SUSY algebra, with $Q$ the supercharge\cite{Witten}. It
interchanges a fermion state and a boson state within a single energy
multiplet called a supermultiplet. The SUSY is the underlying symmetry of
the QH effect in silicene just as in graphene\cite{EzawaPLA}.

\begin{figure}[t]
\begin{center}
\includegraphics[width=0.45\textwidth]{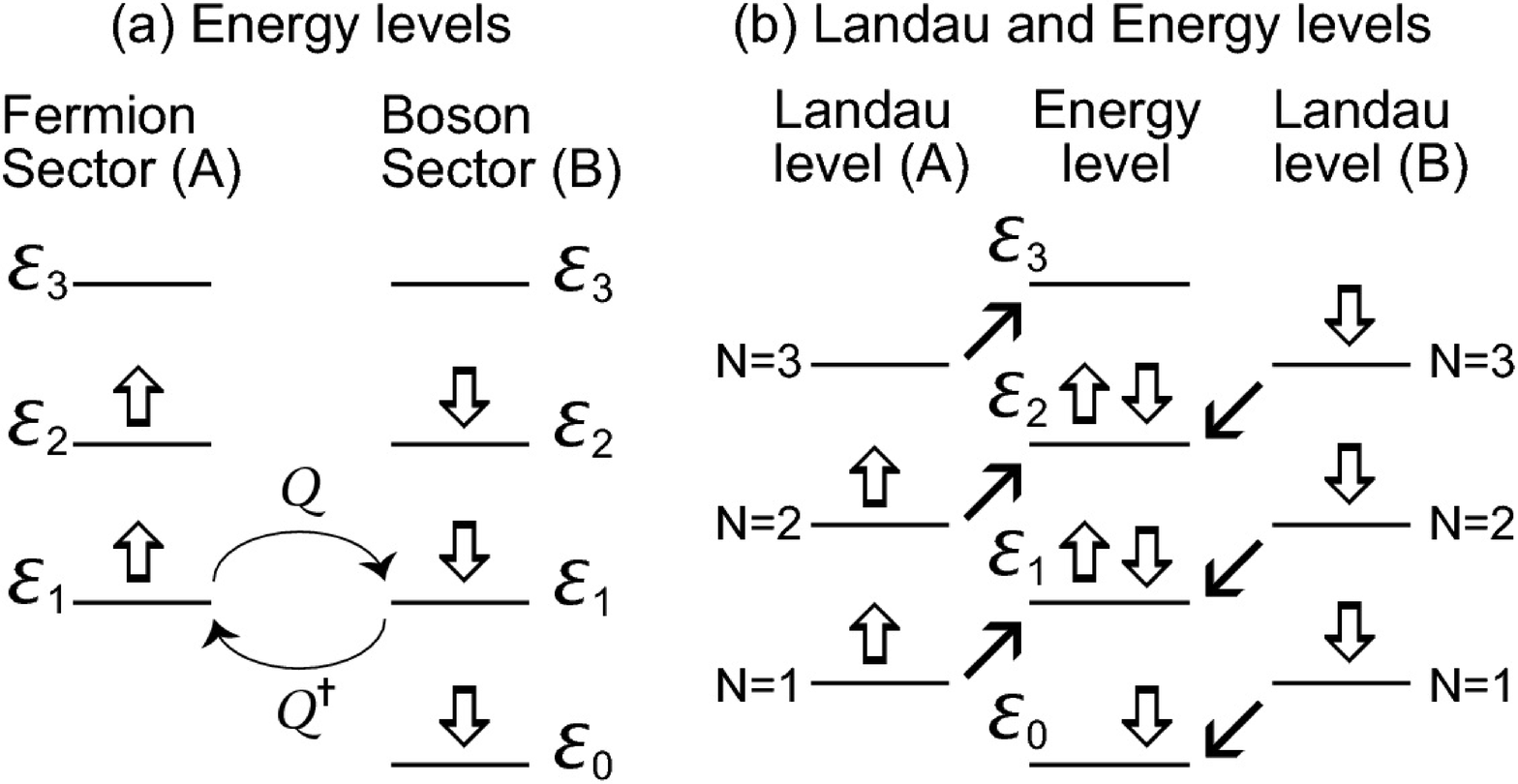}
\end{center}
\caption{{}(a) The illustaration of the SUSY structure. The energy spectra
of the fermion sector (A site) and the boson sector (B site) are related by
the supercharge $Q$. The two states on the same horizontal line have the
same energy, making a supermultiplet, except for the ground state ($\protect%
\varepsilon _{0}$) in the boson sector. (b) The mechanism of the
Landau-level mixing. It is illustrated how a Dirac electron with the energy $%
\protect\varepsilon _{N}$ comes from the $N$th Landau level of the A site
and also from the ($N+1$)th Landau level of the B site in the Dirac K
valley. In this figure electrons are filled up to the level $\protect%
\varepsilon _{2}$. }
\label{FigSiliceneSUSY}
\end{figure}

To construct the fermion and boson sectors explicitly we analyze the
Hamiltonian (\ref{SUSY-Hamil}) more in detail by rewriting it as%
\begin{equation}
H_{\text{P}}^{+}=\hat{a}^{\dag }\hat{a}+\hat{c}^{\dag }\hat{c},
\end{equation}%
where%
\begin{equation}
\hat{c}=\left( 
\begin{array}{cc}
0 & 0 \\ 
1 & 0%
\end{array}%
\right) ,\qquad \hat{c}^{\dag }=\left( 
\begin{array}{cc}
0 & 1 \\ 
0 & 0%
\end{array}%
\right) .
\end{equation}%
Since $\{\hat{c},\hat{c}\}=0$, and $\{\hat{c},\hat{c}^{\dag }\}=1$, $\hat{c}$
and $\hat{c}^{\dag }$ are the fermion annihilation and creation operators
with the ground state $|0\}_{B}=(0,1)^{t}$ and the one-fermion state $%
|0\}_{A}=(1,0)^{t}$, 
\begin{equation}
\hat{c}|0\}_{B}=0,\quad \hat{c}^{\dag }|0\}_{B}=|0\}_{A},\quad \hat{c}%
|0\}_{A}=|0\}_{B}.
\end{equation}%
The boson sector $|N\}_{B}$ and the fermion sector $|N\}_{A}$ are
constructed by operating the boson creation operator $\hat{a}^{\dag }$ to
the ground states in each sectors,%
\begin{align}
|N\}_{A}=& \frac{1}{\sqrt{N!}}\hat{a}^{\dagger N}|0\}_{A}=\left( 
\begin{array}{c}
|N\rangle \\ 
0%
\end{array}%
\right) , \\
|N\}_{B}=& \frac{1}{\sqrt{N!}}\hat{a}^{\dagger N}|0\}_{B}=\left( 
\begin{array}{c}
0 \\ 
|N\rangle%
\end{array}%
\right) .
\end{align}%
The energy of the states is%
\begin{align}
H_{\text{P}}^{+}|N\}_{A}=& (N+1)|N\}_{A}, \\
H_{\text{P}}^{+}|N\}_{B}=& N|N\}_{B}.
\end{align}%
The two sectors are interchanged by the supercharge,%
\begin{align}
Q|N\}_{A}=& \hat{a}^{\dagger }\hat{c}|N\}_{A}=\sqrt{N+1}|N+1\}_{B}, \\
Q^{\dagger }|N+1\}_{B}=& \hat{a}\hat{c}^{\dagger }|N+1\}_{B}=\sqrt{N+1}%
|N\}_{A}.
\end{align}%
The two states $|N\}_{A}$ and $|N+1\}_{B}$ make a supermultiplet. We may
identify $|N\}_{A}$ as the $N$th Landau level at the A site, and $|N+1\}_{B}$
as the ($N+1$)th Landau level at the B site. They are degenerate, as
illustrated in Fig.\ref{FigSiliceneSUSY}(b). An eigenstate of the
Hamiltonian $H_{+}^{\uparrow }$ is a coherent superposition of them as in (%
\ref{TypeUpSpin}). Similarly we can analyze the Hamiltonian $%
H_{+}^{\downarrow }$.

The physical reason of the degeneracy is understood as follows\cite%
{EzawaPLA,EzawaJPSJ}. We rewrite (\ref{PauliHamilX}) in the following form,

\begin{equation}
(H_{\eta }^{s_{z}})^{2}=v_{\text{F}}^{2}\left[ \left( -i\hbar \nabla +e%
\mathbf{A}\right) ^{2}+\eta e\hbar B\tau _{3}\right] +\Delta _{\eta
s_{z}}^{2}\left( E_{z}\right) .  \label{G-PauliHamil}
\end{equation}%
This is essentially the Pauli Hamiltonian. The first term is the kinetic
term, while the second term is the Zeeman term creating the energy gap
between the A site and B site. The Landau level is generated by electrons
making cyclotron motion, which yields the energy $(N+\frac{1}{2})(\hbar
\omega _{c})^{2}$ in the $N$th Landau level. On the other hand, there exists
the intrinsic Zeeman energy $\pm \frac{1}{2}(\hbar \omega _{c})^{2}$ between
the A and B sublattices\cite{EzawaJPSJ}, and their sum is either $N(\hbar
\omega _{c})^{2}$ or $(N+1)(\hbar \omega _{c})^{2}$. Consequently, the
energy of massive Dirac electrons is either $\pm \varepsilon _{N}$ or $\pm
\varepsilon _{N+1}$, as given by (\ref{EigenZeroN}). Namely, Dirac electrons
with the energy $\varepsilon _{N}$ come from the $N$th Landau level of the A
site and also from the ($N+1$)th Landau level of the B site in the Dirac K
valley, as illustrated in Fig.\ref{FigSiliceneSUSY}(b).

Finally we discuss the effect of $\lambda _{\text{R}}\neq 0$. The
Hamiltonian is no longer block diagonal, where off-diagonal block elements
mix up-spin states of $H_{+}^{\uparrow }$ and down-spin states of $%
H_{+}^{\downarrow }$. The mixing occurs solely within the four states
described by (\ref{EigenStateS}). It is enough to consider a mixing between
positive-energy states labelled by [($N+1$)B$\uparrow ,$(N-1)A$\uparrow $]
and [$N$B$\downarrow ,$($N-1$)A$\downarrow $]: See Fig.\ref{FigCross} for
the instance of $N=1$. They have the energy $\varepsilon _{N}^{\uparrow
}=\varepsilon _{N+1}$ and $\varepsilon _{N}^{\downarrow }=\varepsilon _{N}$,
respectively, as we have shown in (\ref{TypeUpSpin}) and (\ref{TypeDownSpin}%
). When $\lambda _{\text{R}}=0$, they cross at $\varepsilon _{N}^{\uparrow
}=\varepsilon _{N}^{\downarrow }$, which implies $\varepsilon
_{N+1}=\varepsilon _{N}$ with (\ref{EigenZeroN}). Solving this equation we
find%
\begin{equation}
E_{z}=\eta E_{\lambda }  \label{AntiCrossPoint}
\end{equation}%
in the K$_{\eta }$ valley, with $E_{\lambda }$ being given by (\ref{CrossE}%
). When $\lambda _{\text{R}}\neq 0$, the Rashba interaction operates between
these two levels. This is a typical quantum mechanical problem. As a result
of the interaction, an anticrossing takes place with the energy gap%
\begin{equation}
\delta \varepsilon =\frac{2\sqrt{2}\hbar a}{\ell _{B}}\lambda _{\text{R}}
\end{equation}%
at the point (\ref{AntiCrossPoint}). There are many level crossings, but
anticrossings occur only at this point.

\section{Conclusions}

We have studied the QH effect in silicene, which is a monolayer of silicon
atoms replacing carbon atoms in graphene. The system is described by a
massive Dirac theory due to a large spin-orbit couplings. We have calculated
the energy spectrum under magnetic field, which exhibits an intriguing
feature as a function of external electric field as in Fig.\ref{FigLLr}. We
have unveiled a rich physical picture behind it. Our finding will be the
first system in which the Landau level is controlled by applying external
electric field parallel to the magnetic field.

I am very much grateful to N. Nagaosa for many fruitful discussions on the
subject. This work was supported in part by Grants-in-Aid for Scientific
Research from the Ministry of Education, Science, Sports and Culture No.
22740196.

\appendix

\section{Tight-Binding Model}

In this appendix we derive the low-energy Dirac Hamiltonian (\ref{DiracHamil}%
) in silicene. The silicene system is described by the four-band
second-nearest-neighbor tight binding model\cite{LiuPRL},

\begin{align}
H& =-t\sum_{\left\langle i,j\right\rangle \alpha }c_{i\alpha }^{\dagger
}c_{j\alpha }+i\frac{\lambda _{\text{SO}}}{3\sqrt{3}}\sum_{\left\langle
\left\langle i,j\right\rangle \right\rangle \alpha \beta }\nu
_{ij}c_{i\alpha }^{\dagger }\sigma _{\alpha \beta }^{z}c_{j\beta }  \notag \\
& -i\frac{2}{3}\lambda _{\text{R}}\sum_{\left\langle \!\left\langle
i,j\right\rangle \!\right\rangle \alpha \beta }\mu _{ij}c_{i\alpha
}^{\dagger }\left( \vec{\sigma}\times \vec{d}_{ij}^{0}\right) _{\alpha \beta
}^{z}c_{j\beta }  \notag \\
& +\ell \sum_{i\alpha }\eta _{i}E_{z}^{i}c_{i\alpha }^{\dagger }c_{i\alpha }.
\label{BasicHamil}
\end{align}%
The first term represents the usual nearest-neighbor hopping on the
honeycomb lattice with the transfer energy $t=1.6$eV, where the sum is taken
over all pairs $\left\langle i,j\right\rangle $ of the nearest-neighboring
sites, and the operator $c_{i\alpha }^{\dagger }$ creates an electron with
spin polarization $\alpha $ at site $i$. The second term represents the
effective SO coupling with $\lambda _{\text{SO}}=3.9$meV, where $\vec{\sigma}%
=(\sigma ^{x},\sigma ^{y},\sigma ^{z})$ is the Pauli matrix of spin, $\nu
_{ij}=\left( \vec{d}_{i}\times \vec{d}_{j}\right) /\left\vert \vec{d}%
_{i}\times \vec{d}_{j}\right\vert $ with $\vec{d}_{i}$ and $\vec{d}_{j}$ the
two nearest bonds connecting the next-nearest neighbors, and the sum is
taken over all pairs $\left\langle \!\left\langle i,j\right\rangle
\!\right\rangle $ of the second-nearest-neighboring sites. The third term
represents the intrinsic Rashba SO coupling with $\lambda _{\text{R}}=0.7$%
meV, where $\mu _{ij}=\pm 1$ for the A (B) site, and $\vec{d}_{ij}^{0}=\vec{d%
}_{ij}/\left\vert \vec{d}_{ij}\right\vert $. The forth term is the staggered
sublattice potential term, where $\eta _{i}=\pm 1$ for the A (B) site. Note
that the first and the second terms constitute the Kane-Mele model\cite%
{KaneMele} proposed to demonstrate the QSH effect in graphene.

Taking the continuum limit is well known except for the Rashba term. With
respect to the Rashba term, we take the six vectors $\vec{d}_{ij}^{0}$ as%
\begin{equation}
\vec{d}_{ij}^{0}=\left( \cos \frac{\pi n}{3},\sin \frac{\pi n}{3}\right) ,
\end{equation}%
with $n=0,1,\cdots ,5.$ Then we obtain%
\begin{align}
\sum_{ij}\left( \vec{\sigma}\times \vec{d}_{ij}^{0}\right) ^{z}& =2ai\sigma
_{x}\cos \frac{ak_{x}}{2}\sin \frac{\sqrt{3}ak_{y}}{2}  \notag \\
& -\frac{2ai}{\sqrt{3}}\sigma _{y}\left( \sin ak_{x}+\sin \frac{ak_{x}}{2}%
\cos \frac{\sqrt{3}ak_{y}}{2}\right) .
\end{align}%
We have taken the K and K' points as%
\begin{equation}
K=\left( \frac{4\pi }{3a},0\right) ,\qquad K^{\prime }=\left( -\frac{4\pi }{%
3a},0\right) .
\end{equation}%
We may expand it as%
\begin{equation}
\sum_{ij}\left( \vec{\sigma}\times \vec{d}_{ij}^{0}\right) ^{z}=\frac{3ai}{2}%
\left( \sigma _{y}k_{x}-\sigma _{x}k_{y}\right) .
\end{equation}%
This is the low-energy continuum version of the Rashba term in the Dirac
Hamiltonian (\ref{DiracHamil}).

\end{document}